\pdfoutput=1

\documentclass[11pt]{article}

\usepackage{acl}

\usepackage{arydshln}
\usepackage{tabularx}
\usepackage{times}
\usepackage{latexsym}
\usepackage{multirow}   
\usepackage{graphicx}   
\usepackage{enumitem}
\usepackage{amsmath}    
\usepackage{amssymb}    
\usepackage{latexsym}
\usepackage[inkscapelatex=false]{svg}
\usepackage{url}    
\usepackage{colortbl, booktabs} 
\usepackage{makecell} 
\usepackage{color}  
\usepackage{array}
\usepackage{hyperref}   
\usepackage[all]{nowidow}   
\usepackage[subtle]{savetrees} 

\usepackage{algorithm}
\usepackage{algorithmic}

\usepackage{microtype}
\usepackage{multicol}
\usepackage{tipa}

\usepackage{inconsolata}

\usepackage[T1]{fontenc}

\usepackage[utf8]{inputenc}

\title{Do as I Say, Not as You Do:\\ Assessing Public Versus Practitioner Perspectives on the Development of Artificial Intelligence and Language Technologies}

\author{Tyler Loakman\textsuperscript{1*, 2},
Chenghua Lin\textsuperscript{2}\\
  \textsuperscript{1}Department of Computer Science, The University of Sheffield, UK \\
  \textsuperscript{2}Department of Computer Science, The University of Manchester, UK\\
    \texttt{chenghua.lin@manchester.ac.uk}}

\begin{document}
\maketitle

\begingroup
\renewcommand{\thefootnote}{*}
\footnotetext{Work performed while at the University of Sheffield.}
\endgroup

\begin{abstract}
  With the public awareness of artificial intelligence (AI) applications such as large language models (LLMs) and generative text-to-video models reaching unprecedented levels, there is an increasing need to reconcile the perspectives held by informed practitioners, who research and develop these technologies, with the interests and opinions of the general public, who constitute the majority of people impacted by technological developments. In this work, we survey practitioners (i.e., academics and industry professionals) alongside members of the general public on perspectives relating to pertinent areas of artificial intelligence, including  perceived benefit and risk, the ability for models to achieve consciousness and output art, and the use of copyright protected material for model training. Alongside comparisons of diverse perspectives on these topics, we additionally contextualise the findings with a discussion of relevant literature that supports and refutes the myriad observed stances.\footnote{\url{https://github.com/tylerL404/do-as-i-say-not-as-you-do}}
\end{abstract}

\section{Introduction}

The advent of technologies such as large language models (LLMs) like ChatGPT \citep{openai2023gpt4} and other artificial intelligence (AI) applications has driven the unprecedented integration of AI into all aspects of life, including healthcare \citep{luo-etal-2025-empec, mondal-etal-2022-global}, education \citep{kumar_ai_education_2025,han-etal-2024-llm,selwyn_future-of-ai-in-education-2022}, entertainment \citep{Hemraj2025-AI-creative-development}, and everyday consumer goods such as fridges and e-cigarettes. The available investment capital for AI technology has additionally resulted in ``AI washing'', where products and services exaggerate the extent and effectiveness of AI integration \citep{Ozturkcan-AI-marketing-2025, Elhajjar2025-ai-washing, Wu-AI-ETF-2022}. As a result, the integration of AI into everyday life causes substantial economic gains for some parties, whilst presenting an economic risk to others \citep{we-wont-be-missed-2025,yuan_Tang_Cao_Wei_Shen_2024}.

There has been a growing body of work in natural language processing (NLP) on the post-training stage of model development, particularly regarding the alignment of model behaviours to human preferences \citep{pan-etal-2025-survey, shu-etal-2025-large}. However, a substantial quantity of human preference data is typically required. One issue that commonly arises during this stage is that such approaches do not adequately account for the plurality of perspectives held by individuals across different cultures and social groups, leaving some demographics under-represented \citep{Zhi-Xuan2025-beyond-preferences, barretto-underrepresented-students}. Similarly, the domain of model debiasing and detoxification relies on prior expectations about what is considered harmful and offensive \citep{Zhi-Xuan2025-beyond-preferences}. However, research shows that over-reliance on such approaches results in the degradation of overall model performance and a preference for abstaining from responding to topics that may be construed as sensitive or contentious, effectively neutering model usability \citep{kaneko-etal-2023-impact, meade-etal-2022-empirical}. 

Alongside the rapid deployment of AI technology and arising challenges in achieving representative alignment for models, there remains a substantial gap in understanding regarding the resulting societal impact. Crucially, the priorities and policies governing those developing AI (frequently driven by chasing leaps in technical capabilities and corporate objectives) may not align with the concerns of the public consumers whose lives these systems fundamentally affect \citep{MetaAIChatbot, AmazonAIOpenLetter, RightToWarnAI}.

To address this critical gap, we conduct a survey to compare the perspectives of AI practitioners (consisting of academics and industry professionals) with the general public. In doing so, we ascertain whether the creators and consumers of such technologies are aligned, thereby gaining a better understanding of the benefits and risks of continued AI development. We summarise the primary contributions of this work as follows:

\begin{itemize}
    \item We conduct a survey of the three primary stakeholders in the development of artificial intelligence: academics, industry professionals, and the general public (dichotomised as practitioners and the public) on topics relating to both practical and philosophical domains (e.g., economic and legal concerns, versus art and consciousness).
    \item Responses are analysed to explore whether practitioners and the public are aligned in their perspectives on AI, assessing whether the development of such technologies is to the benefit of society as a whole.
    \item We provide an overview of the literature surrounding AI development, including areas of legal, economic, and social concern.
\end{itemize}

\section{Related Works} 

\paragraph{Public Perception of AI.}
Existing works have explored the perspectives of different stakeholders in AI development. For instance, \citet{roundtree-public-2024} presents a meta analysis of literature surrounding public perceptions of AI development alongside a sentiment analysis of social media communication on Quora and Reddit. Their findings demonstrate concerns regarding the impact of AI development on employment and the risks of AI driven healthcare. Furthermore, \citet{brauner-public-2023} surveyed 122 members of the German public on opinions regarding personal, economic, industrial, social, cultural, and health-related factors relating to AI development. They identified cybersecurity risks as being the most critical point of public concern when accounting for the interaction between risk and likelihood, whilst threats to careers and personal lives were seen as unlikely, and the position of AI performing unpleasant and undesirable human tasks was viewed as both likely and desirable. Furthermore, \citet{Selwyn2022} analysed the understanding of the Australian public on AI in 2020, prior to the widespread adoption of performant LLMs. Their results showed that most participants had a sufficient working understanding of the technology, even if they self-identified as lacking knowledge.

\paragraph{Comparative Studies.}
Comparing the general public and practitioners, \citet{Pew2025AIExperts} directly surveyed 5k+ members of their American Trends Panel and 1k+ authors of AI publications in 2024 (all of which reside in the United States).\footnote{As a result, this work was both American-centric, and also surveyed members of a questionnaire panel, resulting in a substantial response bias. In contrast, this work is international (though predominantly European), and anonymous.} Results showed that AI experts were significantly more optimistic than the public regarding the potential impact of AI, particularly towards medical care, the economy, and education. We compare the findings of the present work with \citet{Pew2025AIExperts} in Appendix \ref{sec:comparison}. Whilst these findings establish a trend of practitioner optimism and public pessimism, existing comparative works have focussed on broad societal applications (e.g., the economy and healthcare), whilst the present work additionally explores philosophically-laden topics such as art, creativity, and consciousness.

\paragraph{Present Work.}
Building upon these findings, we address gaps in the literature by surveying public and practitioner views on AI with a predominant (though non-exclusive) focus on language technologies. In addition to topics pertaining to practical areas such as the economy and healthcare, we further investigate perspectives surrounding philosophical concerns, such as the ability for AI to generate art and demonstrate consciousness. Such exploration enables a deeper understanding of people's fundamental perspectives regarding subjective topics where opinions are more resistant to change, in comparison to concerns that can be empirically verified (e.g., the impact of AI on economic growth and employment).

\section{Methodology}
\label{sec:methodology}

\paragraph{Recruitment of Participants.} Recruitment was performed through addressing email lists at the lead author's institution, posting on social media, and approaching industry contacts for possible survey dissemination. A wider range of general public respondents was sought through posts on Reddit and the placement of the survey on SurveySwap. All participants were encouraged to share the survey to further eligible participants. Responses were collected in 2025.

\paragraph{Survey Phrasing/Pilot.}
A small scale pilot study was performed to elicit feedback regarding question clarity and logistics (2 practitioners and 2 members of the public). Phrasing was altered for clarity in light of this feedback.

\paragraph{Response Filtering.}
All participants were asked to respond to questions outlining their understanding and familiarity with AI technologies in addition to their understanding of the technical underpinnings of machine learning (e.g., architectures and training methodologies). Participants additionally self-identified their position within AI development as being in academia (e.g., researchers and lecturers in natural language processing, computer vision, and general machine learning), industry (e.g., private researchers and applied AI software developers) or the general public. Using responses to a five-point Likert scale, we removed self-identified in-domain academics and industry practitioners who did not identify their own general and technical AI knowledge to be above or equal to 3. As a result, we filter out mistaken self-identifications (as practitioners necessarily must have sufficient domain-knowledge), whilst not being overly restrictive (e.g., requiring technical knowledge ratings of 4-5), therefore allowing multidisciplinary practitioners who may regularly use AI systems as the focus of their work, but for whom a deep technical knowledge may not be mandatory. Finally, response timestamps were checked for suspicious activity such a botting, of which no evidence was found.

\paragraph{Demographic Information.}
\label{sec:demographics}
In total, the survey received 336 responses: 69 academics, 33 industry professionals, and 234 members of public. Following filtering, this resulted in 76 practitioners across academia and industry, and 212 members of the general public. Over half of respondents were Millennials (born 1981-1996), with similar numbers of Gen X (1965-1980) and Gen Z (1997-2006), at $\approx$20\% each).\footnote{Respondents were required to be at least 18 years of age.}.\footnote{We use the textual generational labels for ease of reference - full date ranges were presented to respondents.} A further small number of respondents were from the Baby Boomer generation (1946-1964). In terms of geographical spread, the majority of respondents were from Europe ($\approx$60-70\%), with a considerable number of North American responses ($\approx$17-30\%). A further nominal number of responses were received from Oceania, East Asia, Southeast Asia, and the Middle East/North Africa (MENA). In total, age and geographical demographics were comparable across practitioners and the public, providing a sample that is not explicitly restricted to a specific locale or other demographic variable. We do not present further results breakdowns by region or age due to the skewed distribution. Additional information regarding the industry/field breakdown of the public respondents is provided in \autoref{fig:public_industry} within Appendix \ref{apx:public}.

\paragraph{Rating Scales.} All non-categorical statements were rated on a 5-point Likert (i.e., ``strong disagree'' to ``strong agree'', or the semantic equivalent for a given question phrasing).\footnote{Participants were only presented with the linguistic labels for the two extremes (i.e., 1 and 5), whilst a middle-point ``neutral'' option is added to figures for clarity.}
Only broad demographic information was collected to encourage honesty, as some respondents may have opinions that challenge the public stance of the organisations they represent. 

Question phrasing has been condensed for inclusion in visualisations, whilst the full original questions are presented in \autoref{tab:stats} within Appendix \ref{apx:questions}, in addition to significance testing for ordinal responses.\footnote{Differences between practitioner and public responses are significant at $\alpha = 0.01$ for all statements.}

\section{Artificial Intelligence Usage}

\begin{figure}
    \centering
    \includegraphics[width=1\linewidth]{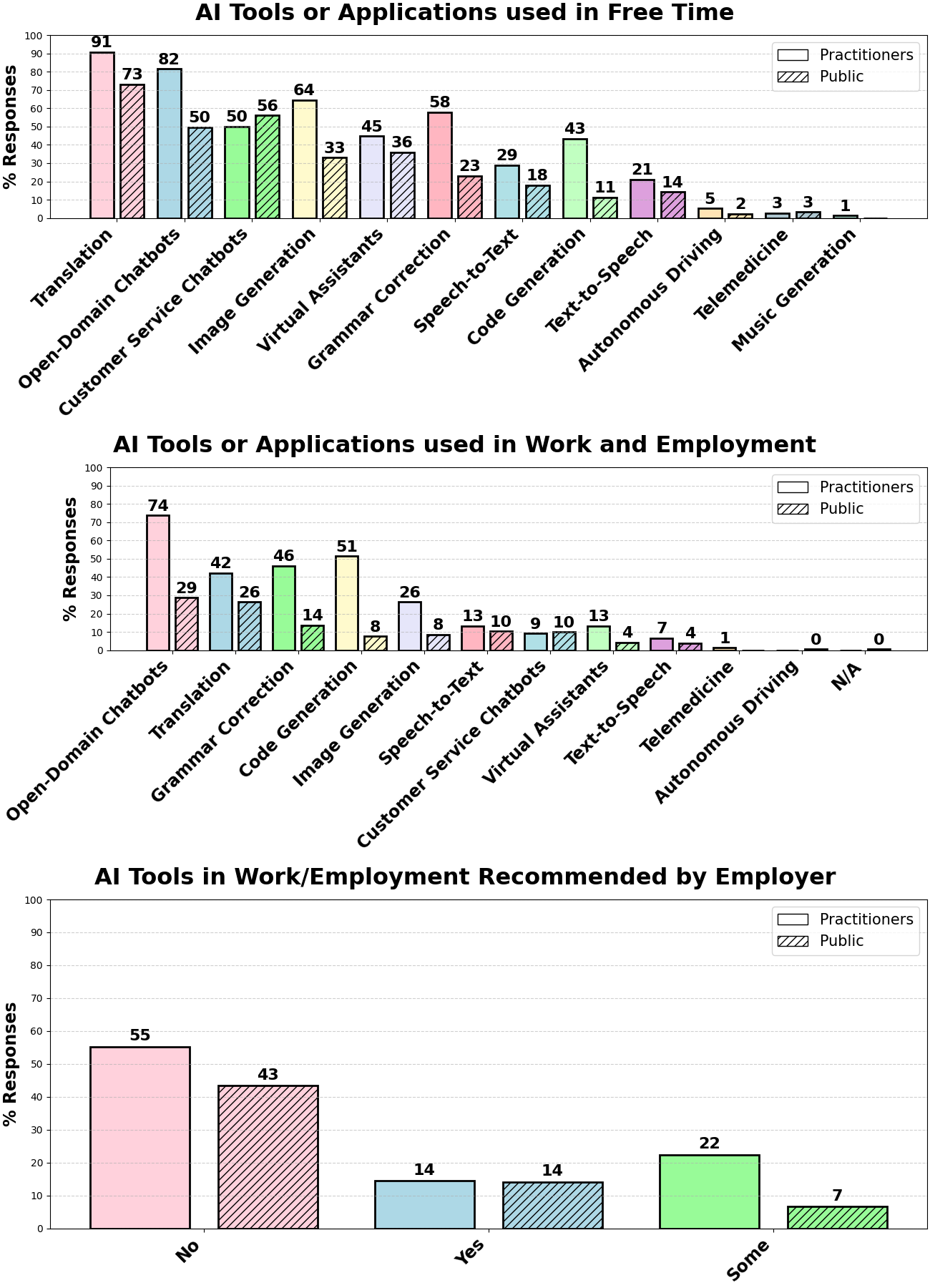}
    \caption{Practitioner and public use of AI technologies in free time and employment.}
    \label{fig:use}
\end{figure}

With the wide range of forms that artificial intelligence and machine learning models take, from simple grammar correction and translation tools to in-line code suggestions and autonomous driving assistance, in \autoref{fig:use} we first outline the different technologies being used both at home and for work.

Similar popularity rankings can be seen for different technologies between practitioners and the public, but with different levels of general use. For instance, the most popular technologies for personal use across both groups are translation tools and open-domain chatbots, with more than half of respondents from both groups using such technologies, whilst a considerable portion of people additionally interact with image generation, virtual assistants, and grammar correction AI ($\approx$20\%). Across both groups, the rarest technologies used by participants are autonomous driving, AI-driven telemedicine, and music generation models - likely owing to their niche application, relatively rudimentary abilities, or high financial barrier to entry. Importantly, over half of public respondents used open-domain chatbots (i.e., LLMs). Such results are unsurprising, with OpenAI revealing that by 2025, approximately 10\% of the world's adult population had interacted with ChatGPT alone \citep{Hitzig2025}.
Regarding AI-usage at work (or in study, for full-time students), less than 30\% of public respondents interact with any, whilst the majority of practitioners make use of open-domain chatbots and specialised code models. Of the respondents that reported using any AI applications in their employment, almost half of the public (43\%) and more than half of practitioners (55\%), did so without endorsement or recommendation by their employer, presenting data protection concerns. Furthermore, 22\% of practitioners and 7\% of the public reported that only a subset of the AI technologies used in employment were endorsed by their employer, while still making use of technologies outside of this scope.

\section{Training Data Collection}

The following section focuses on concerns surrounding the use of publicly available and potentially copyright protected material for model training. Owing to the need for large volumes of data to imbue models with sufficient world-knowledge and enable reasoning abilities, much of the available training data is sourced via scraping large volumes (i.e., the majority) of available web content, to then partially compete against human authors and creators in the same domains. As a result, popular pre-training corpora include collections such as CommonCrawl \citep{CommonCrawl} and RedPajama \citep{weber2024redpajama}, which consist largely of publicly available data.

However, whilst model exist that have been trained on highly curated and ethically collected data, such as Olmo \citep{Olmo3Report,olmo20252olmo2furious,groeneveld-etal-2024-olmo}, which is trained on the Dolma dataset \citep{dolma}, the majority of large-scale pre-training still concerns mass-collected data where little regard is given to the licenses of the original works and the rights of copyright holders.

Whilst at the time of publication, the legal status of AI training remains unclear, there have been a growing number of legal challenges on the topic of model training, including for language generation \citep{AdvanceLocalMedia_v_Cohere_2025}, code generation \citep{doe_v_github_2022}, and image generation \citep{Andersen_v_StabilityAI_2024}. In part, such cases concern the disregard of traditionally upheld copyright protections and the debated matter of model training constituting fair use. This has resulted in loopholes being exploited by some companies, such as Anthropic legally obtaining non-pirated physical copies of print publications and destructively scanning them \citep{TEILMANNLOCK2025106190, bartz_v_anthropic}, following fair-use precedent set by Google Books \citep{AuthorsGuildvGoogle2015}.

Additional complications arise from unclear agreements between platforms and users, where user consent is purportedly given for data to be used for AI purposes, whilst the layperson user is not fully aware of their agreements. Such patterns have been previously observed with the use of end user license agreements (EULAs), where details are obfuscated behind legalese or buried with extensive terms and conditions \citep{CORBETT2019453,Hutton2017}.

\begin{figure}[h]
    \centering
    \includegraphics[width=1\linewidth]{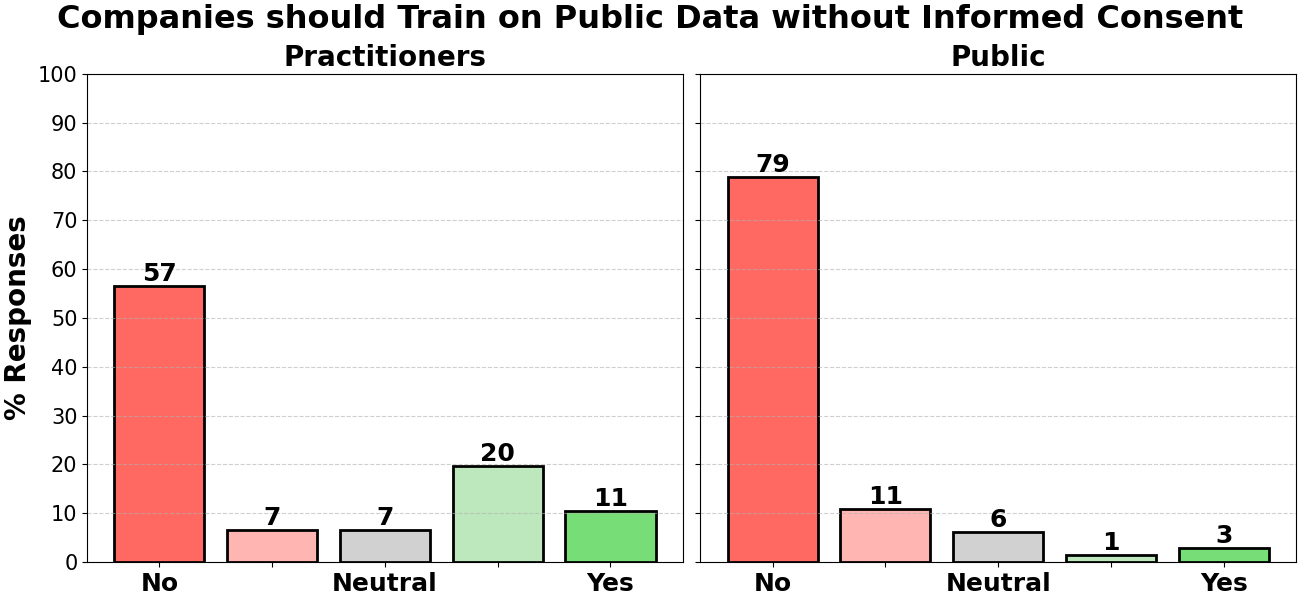}
    \caption{Practitioner and public beliefs concerning AI models being trained on public data without explicit informed consent.}
    \label{fig:copyright}
\end{figure}

\autoref{fig:copyright} presents survey results regarding whether companies should be permitted to train on public data without the explicit consent of copyright holders. Overall consensus of both groups is against the motion that copyrighted material should be without explicit consent (64\% for practitioners and 90\% for the public). However, a clear discrepancy is seen between the groups, with 31\% of practitioners being in favour of being able to make use of copyright protected material without consent, versus only 4\% of the public.

One potential explanation is practitioners viewing the often lengthy and expensive process of obtaining license information and contacting rights-holders to be burdensome, whilst members of the public may not believe that the benefit they receive from more performant AI products outweighs the implications of private organisations profiting from models trained on the work on uncompensated individuals.
In particular, this topic raises a potential conflict of interest in practitioners, with many positions being highly lucrative compared to most employment performed by the public \citep{BONE2025124042,PwC_AIJobsBarometer2025}. As a result, being able to more quickly perform research and develop successful products may lead to increased compensation or career progression for practitioners, whilst from the public perspective many jobs are placed under growing threat of replacement \citep{we-wont-be-missed-2025}. On the other hand, mandating that companies have the requisite permissions for all training data can be viewed as a competitive disadvantage in the AI race, allowing companies in jurisdictions with more relaxed intellectual property laws to have a significant advantage by having access to substantially larger volumes of data \citep{Rosati_2025,Rosati03072019}. We speculate that open-access models with lower barriers to entry (i.e., no fees) are more likely to receive licensing permissions owing to the wider democratisation of their capabilities.

\section{Art \& Creativity}

Whilst AI applications have previously been developed predominantly for more mundane or supra-human tasks such as text summarisation and cancer screening, models have now become capable of generating articulate, engaging text and realistic video, resulting in such systems increasingly being used in tasks that have historically relied on human creativity, such as writing narratives \citep{teleki-etal-2025-survey} and poetry \citep{goncalo-oliveira-2017-survey} or designing infographics \citep{ghosh-etal-2025-infogen,ZHU202024} and movie trailers \citep{Yao2025,Argaw_2024_CVPR}. 

\begin{figure}[h]
    \centering
    \includegraphics[width=1\linewidth]{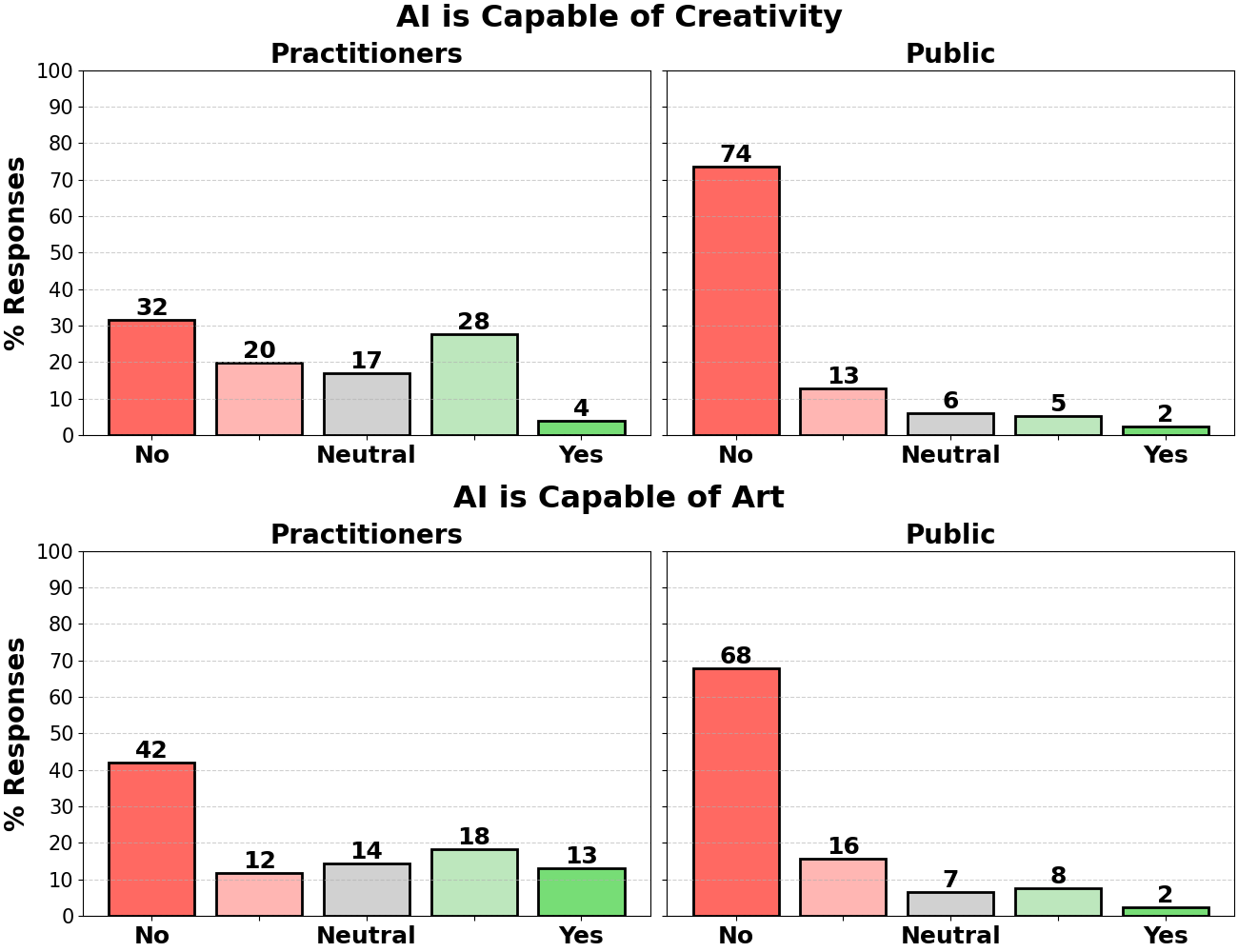}
    \caption{Practitioner and public beliefs concerning the ability of AI models to be creative and generate art.}
    \label{fig:creativity}
\end{figure}

Consequently, respondents were surveyed on their perspectives regarding the ability for AI models to demonstrate creativity and generate art - two philosophically-loaded concepts that are frequently related to intention and embodied experience, which AI may be considered to lack \citep{Leong2025Ethical}. Results are presented in \autoref{fig:creativity}. Across both practitioners and the public, consensus is against AI models being capable of either of these feats. However, responses within practitioners on the topic of AI creativity are less unanimous, with only a small majority (32\%) being strongly of the belief that such systems are incapable of creativity, compared to  74\% of public respondents.

This discrepancy between creativity and art highlights a key insight into the differing perspectives of both parties. Whilst art is often seen as a highly human-centric concept that requires intention and meaning, creativity may also be defined as the ability to produce unique ideas \citep{gayle-creativity-2017}. Consequently, technologies such as large language models are capable of rearranging sequences observed from training data to output novel linguistic strings (and therefore ideas). Practitioners may be more likely to consider such outputs as being creative, even if the ideas are amalgamations and abstractions of existing human thoughts contained in training corpora. Such comparatively shallow abilities have been referred to as pseudo-creativity \citep{RUNCO2023100063}. Even within humans, there is debate as to whether creativity is truly novel, or the result of exposure to embodied experiences that constitutes our training data \citep{lynn-human-learning}.

Extensive work has demonstrated that people are more likely to rate content negatively when it is suggested that it has been created by AI rather than another human, regardless of tangible characteristics \cite{Bellaiche2023}. This preference for human-created content over machine created content is a simple an extension of longstanding patterns observed alongside industrialisation, such as the value placed behind handmade items versus the perceived coldness of machine produced products \citep{degirmenci-2025-handmade, Fan2024, fuchs-handmade-2015, kruger2004effort}. The implications for future development are myriad, primarily concerning the nature of human-in-the-loop (HITL) generative systems and the extent to which human intention and intervention is necessary for the output of a human-in-the-loop system to retain deeper meaning as it relates to art and creativity. For instance, whilst such systems may be frame as collaboration between humans and AI, determining the level of agency and ownership of either participant remains difficult \cite{carrera-egg}, particularly for creative domains \cite{info16100909, loakman-etal-2025-whos}.

\section{Thought \& Consciousness}

Whilst once firmly in the remit of science fiction \citep{Nader2024, Hermann2023, Hudson2023}, the concept of artificial consciousness and the ability for machines to emulate internal cognitive processes such as thought is progressively being viewed as more viable. A former Google engineer made news by claiming that the LaMDA \citep{laMDA} language model was sentient \citep{Tiku2022Google}. Whilst dismissed by the community at large, this presented one of the first notable moments where an in-domain practitioners with expertise made positive claims regarding the sentience of computational models. Previously, such claims were laregely limited to non-experts whose beliefs about sentient machines may have been dismissed as a lack of understanding \citep{Selwyn2022}.

Implications for the consciousness of such technologies includes needing to ensure that such models are not subject to undue suffering and other protections analogous to human-rights \cite{metzinger-suffering-2021, schwitzgebel-ai-rights,laukyte-personhood-2019}. Furthermore, the consciousness of such models has further potential legal implications pertaining to the allocation of blame and responsibility for the outputs of such technologies such as harmful language \citep{Veluwenkamp2025, Noh2025}.

\begin{figure}[h]
    \centering
    \includegraphics[width=1\linewidth]{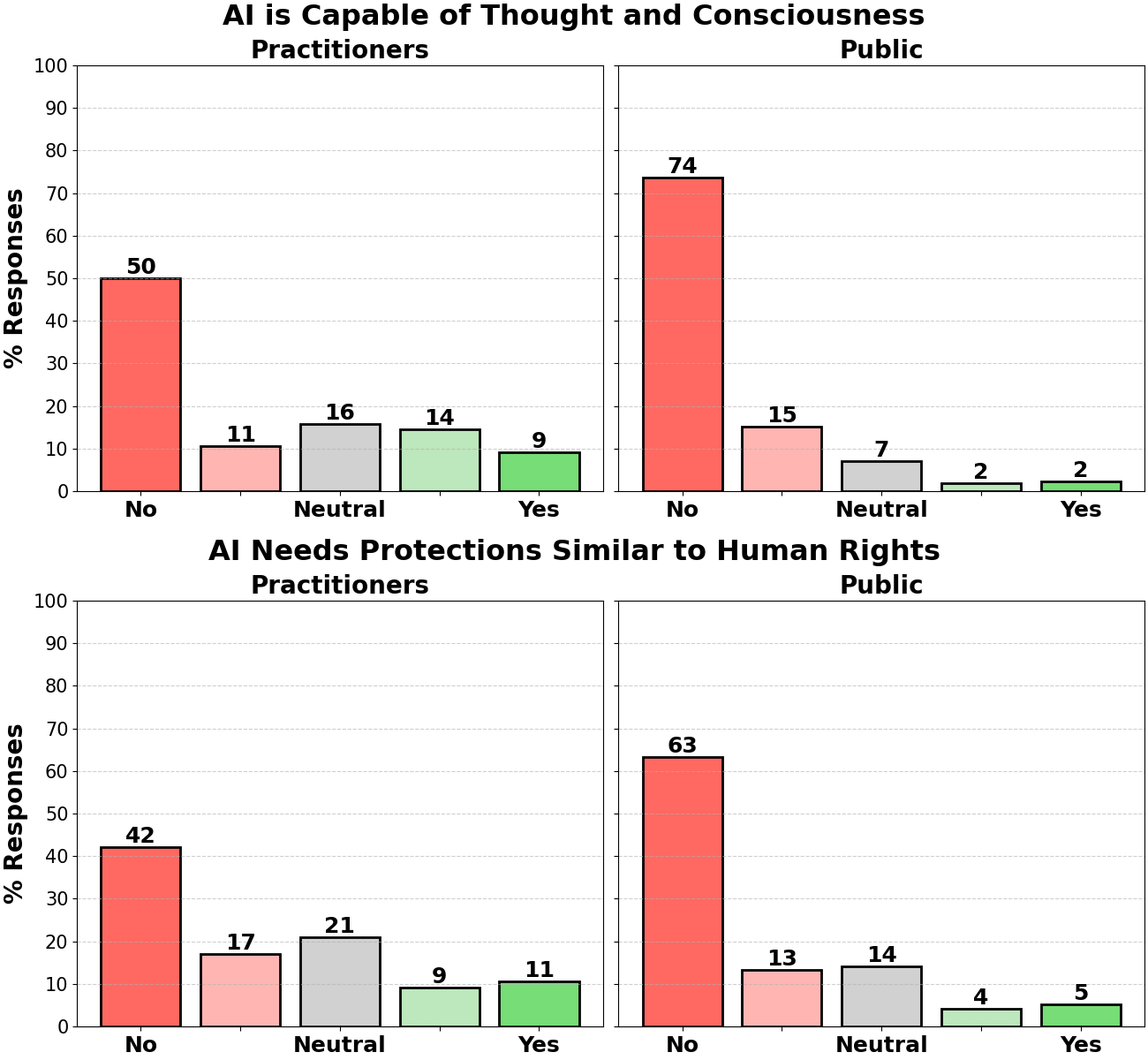}
    \caption{Practitioner and public beliefs concerning the ability for AI systems to be capable of thought and consciousness, and the likelihood of requiring legal protections.}
    \label{fig:consciousness}
\end{figure}

Survey respondents were queried on whether AI models are, or will become, capable of thought and consciousness, and the consequent real-world implications regarding legal protections for AI analogous to human rights. The results in \autoref{fig:consciousness} demonstrate that by and large, both practitioners and the public dismiss the potential sentience of such models, with approximately half of all respondents strongly rejecting both claims. However, only 42\% of practitioners strongly rejected the claim that AI models will ever require human-rights style protections, whilst 17\% and 21\% gave a weaker rejection or were neutral on the topic, respectively. Furthermore, for both questions and for both practitioners and the public, we see at least some respondents giving positive responses, demonstrating that the belief in artificial consciousness and the subsequent need for legal protections is not an entirely fringe belief.

\section{Societal Impact}

\begin{figure}[h]
    \centering
    \includegraphics[width=1\linewidth]{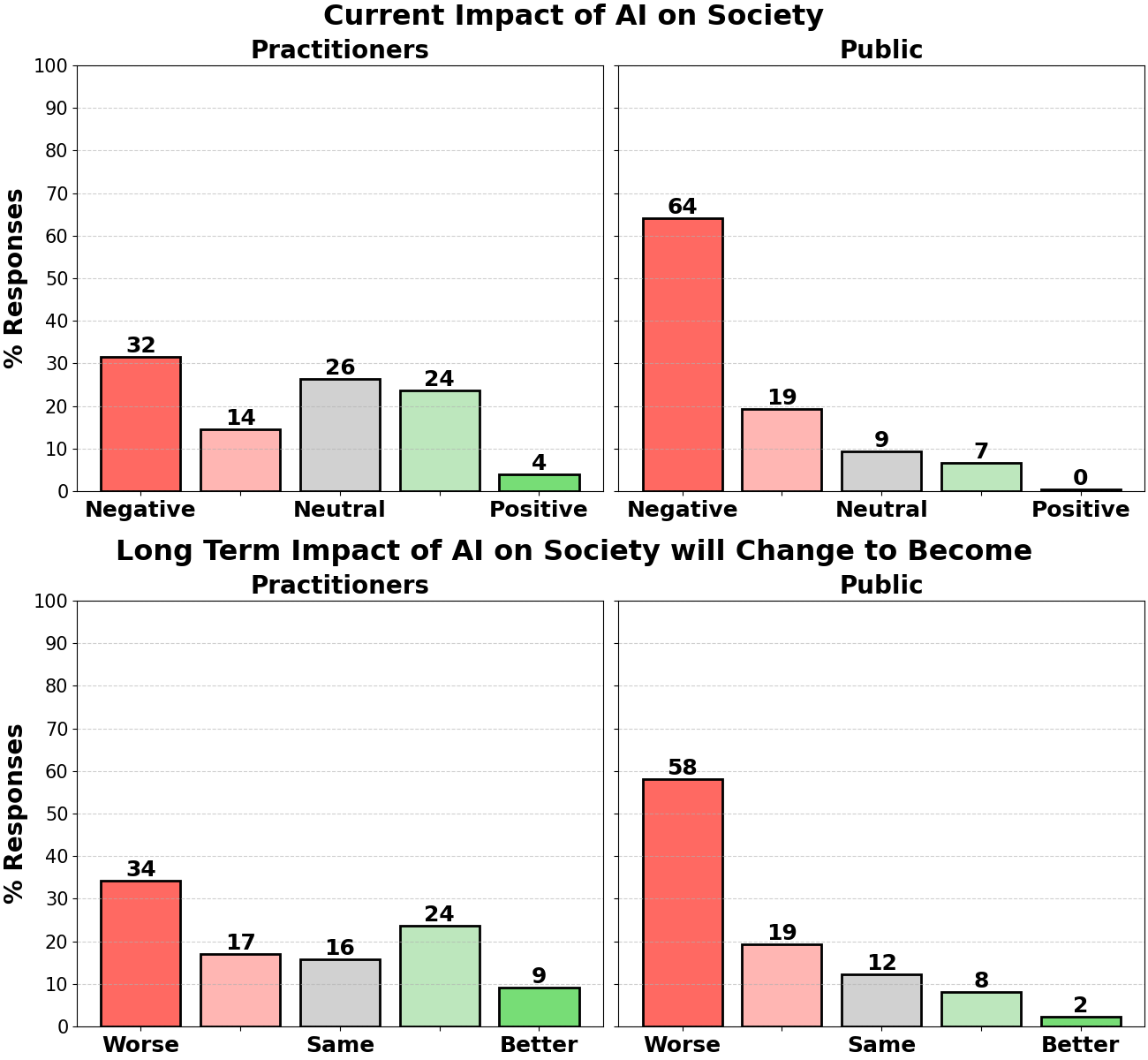}
    \caption{Practitioner and public beliefs concerning the present and future societal impact of AI.}
    \label{fig:impact}
\end{figure}

In this final section, we investigate the aspects of society that people believe are impacted by AI in a net positive or negative way.
\autoref{fig:impact} presents public and practitioner perspectives on the overall impact of AI systems on society at large, both now and in the future. The results show that the public have a much more negative view of the societal impact of AI, with a strong majority of respondents having negative views on the impact of AI at the time of response (83\%), and longer term forecasts (77\%), versus less than half of practitioners (46\% and 51\%, for the present and future, respectively).

\begin{figure}[h]
    \centering
    \includegraphics[width=1\linewidth]{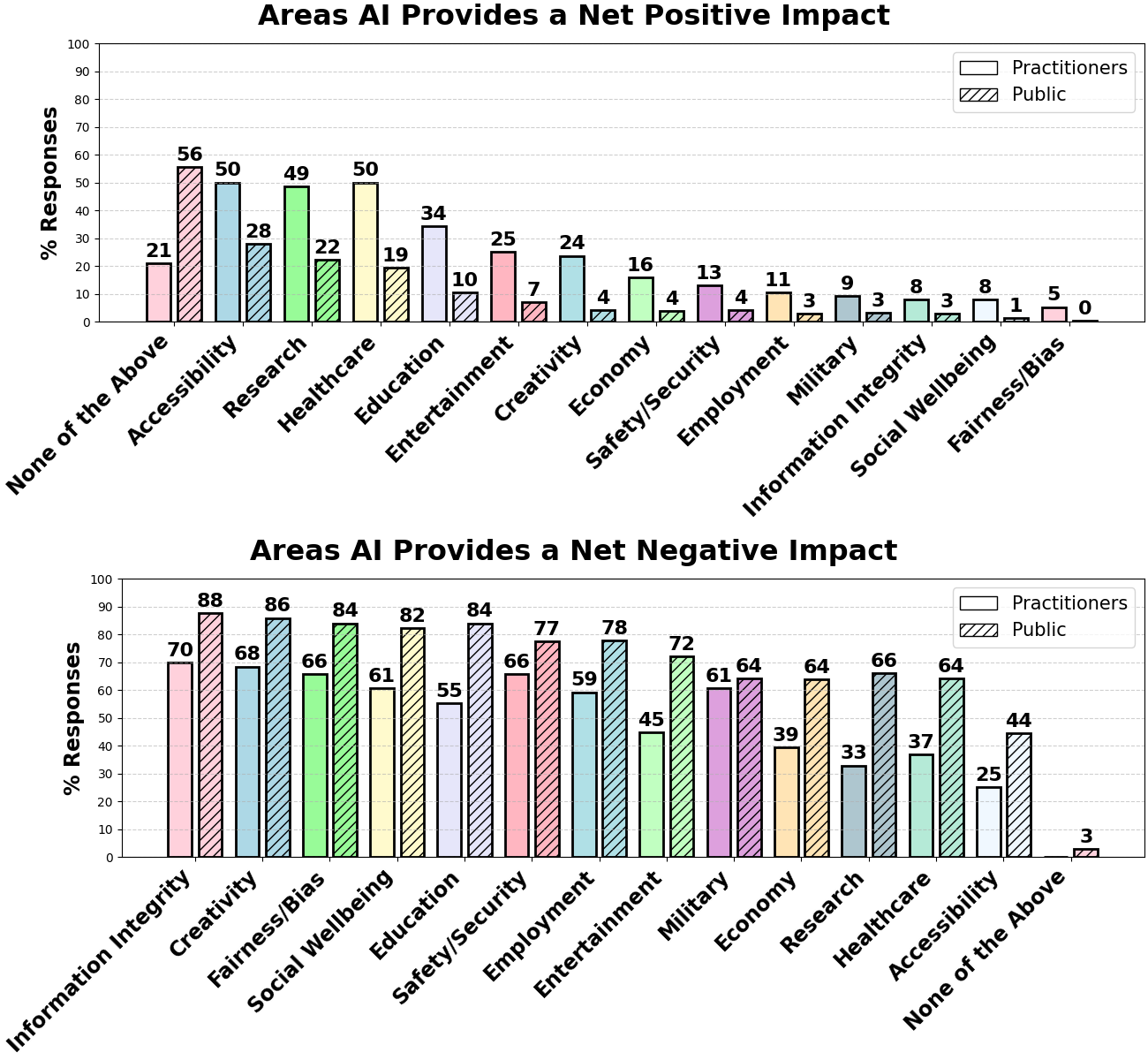}
    \caption{Practitioner and public beliefs concerning the subdisciplines of AI that present overall positive or negative impacts on society.}
    \label{fig:impact-areas}
\end{figure}

\autoref{fig:impact-areas} presents a breakdown of the specific domains and industries where AI is considered to be a net positive or negative. Most strikingly, 56\% of public respondents and 21\% of practitioners believe that AI provides no net benefit to any of the presented areas. In contrast, 3\% of public respondents take the optimistic stance that none of the proposed areas are impacted by AI in a net negative way. Whist raw percentages differ, the results for both net positive and net negative areas are comparable between both practitioners and the general public, demonstrating that specific areas of concern are shared across the surveyed population.

Of the most positively viewed domains, disagreement between the public and practitioners is highest for research, healthcare, education, entertainment, and creativity - with an approximately 2-4x higher proportion of practitioners presenting positive views of the potential impact of AI in these domains than public respondents. 

Overall, the finding that half of surveyed practitioners view AI's societal impact to be net negative is concerning. One interpretation of this finding is that practitioners with this view may wish to remain involved in the field in order to help direct AI development towards positive societal utility, even if they view the domain as a whole as trending negatively. In the following section, we elaborate on some of the key areas of societal impact:

\subsection{Accessibility}
AI technologies can significantly lower barriers to partaking in activities that have traditionally required specialised skill sets and knowledge \citep{app14188236,Dessimoz2024}. Whilst the dawn of the internet presented a positive step in this direction, much specialised knowledge remained complex and difficult to understand. Consequently, the use of technologies such as large language models to summarise and explain concepts in a more accessible, simplified manner further reduces barriers. Barriers to other skills have likewise progressively been reduced, such as being able to create music or animations. However, reducing the skill level required to partake in activities has been shown to reduce a skill's perceived value, therefore garnering negativity from those who learnt with less assistance \citep{kruger2004effort}. Additionally, the over-reliance on AI aids may generally reduce the ability levels of a population over time \citep{soc15010006,Macnamara2024}. Alongside lowering the overall barrier to entry for the general public, AI technologies also provide assistance to those with disabilities, for whom access to a given activity may have otherwise been impossible, rather than just requiring more effort \citep{Brotosaputro2024AIPowered}.

\subsection{Research}
The use of AI in academic research has enabled the significant scaling of experiments, resulting in a faster rate of progress \citep{nolte-ai-driven-experimental-design,gottweis2025aicoscientist, lu2024aiscientist}. However, in specific research domains, the use of AI has increasingly presented challenges. For instance, the pre-print platform arXiv decided to halt the acceptance of survey papers and opinion-pieces due to an influx of low-quality AI generated works \citep{arXiv2025Attention}, whilst many academic venues report the use of AI by reviewers, usually without disclosure. This results in a pipeline where research works have their aims and hypotheses generated by AI, the methodology implemented with AI, and the results written into papers by AI, to be later reviewed by AI \citep{gottweis2025aicoscientist,lu2024aiscientist}. This is compounded by the volume of research works published in fields such as machine learning and natural sciences \citep{van-der-zwaard-articles-impact-2020,pontis-2017-up-to-date,bornmann-growth-rates-2015}, meaning that a significant number of these works go untouched by human eyes throughout their entire life cycle.

\subsection{Healthcare}
The survey responses show that the impact of AI on healthcare is considered positive by many people. For areas such as genome sequencing, early screening of life threatening conditions, drug discovery, and other big-data-reliant healthcare fields, AI and machine learning are integral to meaningful, timely progress \citep{Bajwa2021-sr, Zhavoronkov2019-dy}. Additionally, the availability of AI doctors may help to alleviate strains on medical systems, including in less economically developed nations where medical expertise may not be readily available at the scale required for the population \citep{Yousef_Schmollgruber_2024}. However, negative sentiment arises when considering the use of AI for final diagnoses and decision-making roles, such as determining the likelihood of survival for someone on life support, without human intervention being required for the final decision.

\subsection{Wellbeing}
In terms of mental health and social wellbeing, there has been an increasing number of individuals turning to AI assistance for therapy and counselling services, owing to the relative affordability and the perceived anonymity of conversing with a language model \citep{cox_ephemerality_2025, lee-disclosure-2020}. However, most implementations of AI for this task rely on third party hosted instances of proprietary models, where sensitive data is shared with international corporations, rather than self-hosted open-source implementations that require substantial financial and educational investments to use. Additionally, model debiasing and detoxification are not solved issues, resulting in instances where models have prompted vulnerable users to either make unsupported medical decisions, or take their own lives whilst in depressive episodes \citep{schoene2025forargumentssakeharm,judd2025independentclinicalevaluationgeneralpurpose, arnaizrodriguez2025helpharmevaluationmental}.

\subsection{Information Integrity}
As the most negatively viewed impact of AI on society in our survey, information integrity refers to the reliability and trustworthiness of outputs presented by such technologies. Technologies such as large language models are able to facilitate bad actors in the creation and propagation of fake or misleading information at an unprecedented scale, largely going undetected by users \citep{williams-election-disinformation-2025, radivojevic2024humanperceptionllmgeneratedtext, chen2024llmgenerated}. Such materials exist both as web pages and as bots acting as real social media users to promote anger and civil unrest \citep{Radivojevic2024LLMsAU}. Further adding to the concern, whilst image and video evidence of an event was once considered almost infallible, developments in the automatic generation of such mediums has slowly eroded the utility of such evidence \citep{Iannuzzi2025,Pataranutaporn-AI-memories-2025,maras-video-authenticity-2019}. Combined with the growing distrust of governments and news media, false information is able to propagate faster than ever \citep{Garry2024-co,Prats2024Lack,Mittelstadt2023, Ognyanova2020Misinformation}. Furthermore, even within the scope of expected use, hallucination remains a prevalent concern, casting doubt on the factual accuracy of model outputs, requiring additional verification that many users skip. As such problems worsen and more content becomes model-generated, progress towards realising the dead internet theory increases \citep{Walter2025, Muzumdar2025DeadInternet}.

\section{Conclusion}

This work contributed further insights into the differences of opinion held by the individuals responsible for the development of artificial intelligence technologies (i.e., practitioners) and the members of public that constitute the majority of people impacted by such developments. Through a survey, we highlight key differences between both parties, particularly concerning opinions on copyright responsibilities, societal impact, and the philosophical concerns of art, creativity, and consciousness. Our results support the findings of prior work, empirically demonstrating that practitioners have more optimistic views toward AI technologies than the general public. We argue that such differences in opinion between practitioners and the public are not irreconcilable, but highlight the need for the public to become more well-informed about the nature of AI technologies, as well as the need for practitioners to engage more actively with stakeholders outside of the technology industry and corporate interests to better understand the perspectives of society at large.

\section*{Limitations}
We performed a survey of opinions on the development of artificial intelligence from the perspectives of practitioners and the general public. However, we were only able to capture a small subset of opinions from each of these parties. Owing to logistics and ethics approval procedures, we were unable to collect responses from individuals in rural communities across the globe with limited access to technology, in addition to the young (who were unable to participate in this work due to the 18+ age restriction), and the the elderly. Due to the survey being presented only in English, we were additionally not able to receive responses from individuals who are not fluent in English.
Whilst we did not enforce a geographical boundary on responses to the survey, the primary demographic represented in this work is that of Millennial Europeans, rather than a stratified sample of the world's true population for representativeness. We highly encourage research groups globally to perform further work in this vain (as well as longitudinal analyses of how perspectives evolve over time as the technology advances). Furthermore, the largest field from which public respondents were recruited from is technology (excl. ML/AI), constituting ~30\% of respondents. However, no statistically significant difference was found between members of the public working in technology and those not.

\section*{Acknowledgments}
Tyler Loakman was supported by the Centre for Doctoral Training in Speech and Language Technologies (SLT) and their Applications funded by UK Research and Innovation [grant number EP/S023062/1].

\bibliography{custom}

\appendix
\section{Appendix}

\subsection{Comparison to Existing Work}
\label{sec:comparison}
In the following section, we compare our findings to those reported in \citet{Pew2025AIExperts}, where comparable lines of questioning are available. Regarding long term impact of AI (\autoref{fig:impact}) we observe less optimistic results, with only 33\% of practitioners and 10\% of the public having favourable views, versus 17\% of the public and 56\% of the experts surveyed by the Pew Research Center.
Furthermore, regarding specific areas of benefit (\autoref{fig:impact-areas}), we observe no specific domain in which the majority of practitioners or the public believe AI will present an overall benefit. In contrast, \citet{Pew2025AIExperts} observe majority positive perspectives towards AI's effect on the economy and medical care by practitioners, whilst the public remain mostly pessimistic.
However, our results echo the overall finding that practitioners are almost universally more optimistic than the public. However, regarding AI concerns, both the present work and \citet{Pew2025AIExperts} demonstrate that inaccurate information (i.e., information integrity) is one of the most pressing concerns.

\subsection{Original Phrasing \& Statistical Testing}
\label{apx:questions}
The exact question phrasing as presented to participants is presented in \autoref{tab:stats}, alongside the results of statistical significance tests regarding differences between practitioner and public responses.

\begin{table*}[h]
\centering\scriptsize
\begin{tabularx}{\textwidth}{cXccccc}
\toprule
\textbf{Figure} &\textbf{Original Question} & \textbf{Practitioner} & \textbf{Public} & \textbf{$U$} & \textbf{$p_{\text{BH}}$} & \textbf{$r$}\\
\midrule
-& How familiar are you with the general idea of AI and ML and their applications? &-&-&-&-&-\\
\hline
-& How familiar are you with technical aspects of AI/ML (e.g., neural networks, deep learning)? &-&-&-&-&-\\
\hline
\ref{fig:use}&Which AI-powered tools or applications have you used in your own free time?&-&-&-&-&-\\
\hline
\ref{fig:use}&Which AI-powered tools or applications have you used to aid in your work/employment?&-&-&-&-&-\\
\hline
\ref{fig:use}&Was using the AI-powered tools selected above in your work/employment mandated/recommended by your employer?&-&-&-&-&-\\
\hline
\ref{fig:copyright}&Do you think that AI/ML companies should be able to train models on publicly available data (e.g., social media posts, PDFs, YouTube videos, etc.) without explicit informed consent? & 2.21 & 1.39 & 10,205.0 & $< 0.001$ & 0.267\\
\hline
\ref{fig:creativity}&Do you believe that AI is capable of generating art? & 2.49 & 1.61 & 10,653.5 & $< 0.001$ & 0.322\\
\hline
\ref{fig:creativity}&Do you think that AI is capable of thinking and being conscious? & 2.22 & 1.44 & 10,407.5 & $< 0.001$ & 0.292\\
\hline
\ref{fig:consciousness}&Do you believe that AI/ML systems are capable of being creative? & 2.53 & 1.50 & 11,738.0 & $< 0.001$ & 0.457\\ 
\hline
\ref{fig:consciousness}&Do you think that an AI system will ever need to be granted protections similar to human-rights? & 2.29 & 1.75 & 9,927.5 & $< 0.001$ & 0.232\\
\hline
\ref{fig:impact}&In its current state, how do you perceive the impact of AI on society?& 2.54 & 1.60 & 11,400.0 & $< 0.001$ & 0.415\\
\hline
\ref{fig:impact}&How do you perceive the impact of AI on society to change long-term (e.g., the next 10 years)?& 2.57 & 1.77 & 10,628.0 & $< 0.001$ & 0.319\\
\hline
\ref{fig:impact-areas} &In which areas do you believe AI provides an overall positive impact?&-&-&-&-&-\\
\hline
\ref{fig:impact-areas} &In which areas do you believe AI provides an overall negative impact?&-&-&-&-&-\\
\bottomrule
\end{tabularx}
\caption{Original questions as posed to survey respondents. Two-sided Mann-Whitney $U$ Test results comparing practitioner and public responses are provided, alongside mean Likert scores and absolute rank-biserial correlation ($r$) for effect size. $p$-values are adjusted using the Benjamini-Hochberg procedure across all 12 evaluation metrics. Practitioner $N = 76$, Public $N = 212$.}
\label{tab:stats}
\end{table*}

\subsection{Additional Demographic Information}
\label{apx:public}

\begin{figure*}
    \centering
    \includegraphics[width=1\linewidth]{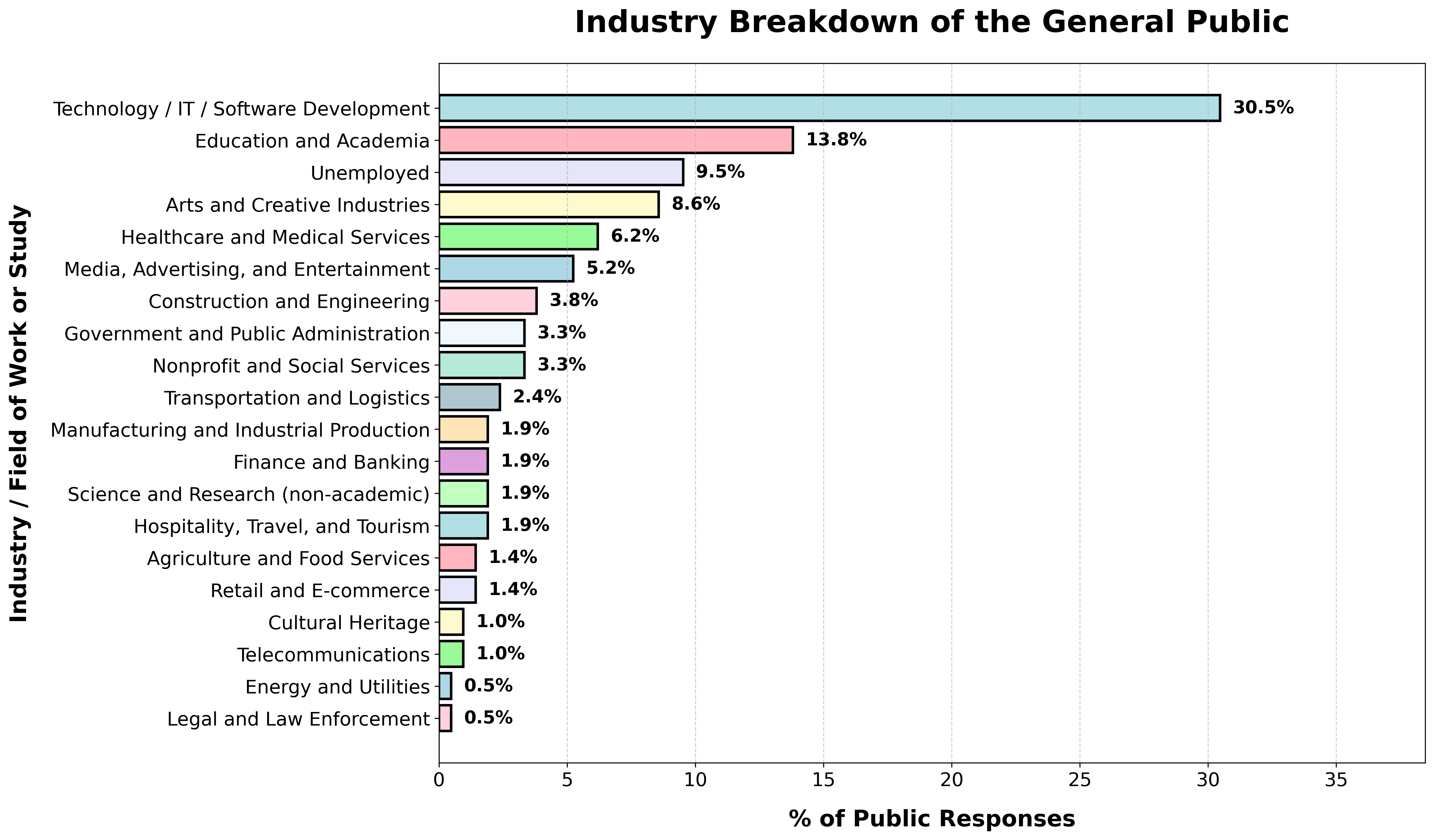}
    \caption{Breakdown of industries within which public respondents operate.}
    \label{fig:public_industry}
\end{figure*}

Additional information regarding the specific fields that public respondents came from is presented in \autoref{fig:public_industry}.

\end{document}